# 多层次的 Android 系统权限控制方法[*]

罗　杨[1,2,3], 张齐勋[1,2,3], 沈晴霓[1,2,3], 刘宏志[1,2,3], 吴中海[1,2,3]

[1](北京大学 软件与微电子学院,北京　100871)
[2](网络与软件安全保障教育部重点实验室(北京大学),北京　100871)
[3](北京大学 服务计算与大数据技术研究室,北京　100871)

通讯作者: 罗杨, E-mail: luoyang@pku.edu.cn, http://www.pku.edu.cn

**摘　要**: 随着Android智能平台的普及,其安全问题日益受到人们关注.在底层安全方面,部分root工具已经实现了对最新版本 Android 的 root 提权,从而给恶意软件滥用权限造成可乘之机;在上层应用安全方面,目前还没有能够在应用权限进行有效管理的方法.基于安全策略的思想,提出了一种 Android 应用权限动态管理机制,利用安全策略对授权进行描述,在 Android 框架层设置权限检查点,并调用请求评估算法进行授权评估,从而实现对应用行为的监控.实验结果表明,该方法能够有效管理 Android 应用权限的正常调用,约束非法调用,并且系统开销较小.
**关键词**: 权限管理;安全策略;SELinux;请求评估;权限检查

## Android Multi-Level System Permission Management Approach

LUO Yang[1,2,3], ZHANG Qi-Xun[1,2,3], SHEN Qing-Ni[1,2,3], LIU Hong-Zhi[1,2,3], WU Zhong-Hai[1,2,3]

[1](School of Software and Microelectronics, Peking University, Beijing 100871, China)
[2](MoE Key Laboratory of Network and Software Assurance (Peking University), Beijing 100871, China)
[3](Research Laboratory of Service Computing and Big Data Technology, Peking University, Beijing 100871, China)

**Abstract**: With the expansion of the market share occupied by the Android platform, security issues (especially application security) have become attention focus of researchers. In fact, the existing methods lack the capabilities to manage application permissions without root privilege. This study proposes a dynamic management mechanism of Android application permissions based on security policies. The paper first describes the permissions by security policies, then implementes permission checking code and request evaluation algorithm in Android framework layer. Experimental results indicate that the presented approach succeeds in permission management of Android applications, and its system overhead is low, which makes it an effective method for Android permission management.

Android 近年来发展迅速,在 2014 年年底其市场占有率已经高达 76%,远超第二名成为最流行的移动终端操作系统[1].由于 Android 在设计之初对系统安全性考虑不周,虽然历经多个版本的漏洞修补和安全增强,其目前仍面临着很多安全威胁[2].总体上主要包括以下几个方面:① 应用程序权限混乱,缺乏管理[3];② 恶意软件吸费、扣费[4];③ 移动平台杀毒软件无法很好地抵御病毒入侵[5];④ 非法 root 容易破坏系统现有安全体系[6].造成这些问题的原因,一方面是由于 Android 平台的漏洞层出不穷,给病毒、木马、恶意软件等以可乘之机;另一方

面,Android 系统结构复杂,权限控制机制也不够健全.Android 底层采用的是 Linux 内核,上层则封装了 Java 虚拟机并在此基础上运行应用,这导致其底层安全和上层应用安全无法采用相同的机制.

Android 系统底层依赖于 Linux 内核及各种 Linux 服务,其权限管理也采用的是 Linux 系统自有的 DAC 模型.DAC 模型由于其访问控制的粗粒度,资源所有者可以任意修改访问权限等问题一直广为诟病,因此 Google 从 Android 4.3 开始,逐渐引入基于 SELinux 安全加固机制,并在 Android 5.0 版本上对所有 domain 执行 enforcing 模式.SELinux 同时包括 Domain Type Enforcement(DTE),Role-Based Access Control(RBAC)和 Multi-Level Security(MLS)这 3 个安全模型的实现.

然而,目前的 SELinux 仍存在一些问题,如 RBAC,MLS 等安全机制实际上并没有启用,Android 实现了 MLS 模型及其验证机制,但是由于 MLS 策略制定需要依据 Bell-LaPadula(BLP)模型为每一个主体和客体设置(密级,范畴集)二元组表示的安全级,过于复杂,因此当前 MLS 模型中只有一个安全级 $s0$,模型也就无法发挥其应有作用.除此之外,SELinux 由于支持策略的动态加载,还存在策略容易被恶意代码动态篡改等威胁.以上诸方面均会导致 Android 系统在面临攻击时的漏洞缓解能力大打折扣.经过实验,结果表明,目前的 SuperSU,KingRoot 等工具已经实现对 Android 5.1.0 R3 版本的 root 提权,从而允许第三方应用获取系统最高权限,给系统带来非常大的安全隐患.

Android 上层平台主要负责对应用权限的管理.由于 Android 对应用权限管理机制的安全不够重视,一直以来缺乏对应用权限进行管理的统一的安全机制.目前学界和业界针对 Android 应用的权限管理提出了很多改进方案,主要包括以下 3 种:重打包、系统补丁和注入钩子.

重打包技术的代表应用是 App Shield,此方法主要通过更改应用程序 APK 安装包中的 AndroidManifest.xml 文件来实现权限的修改,然后再重新打包[7].本方法技术上易于实现,但是重新打包势必会破坏应用原有的签名,从而导致应用安装时显示的厂商信息出现错误,给用户带来误导.

系统补丁方式的典型应用就是 PDroid[8],此方法通过对内核进行修改,可以有效地在系统 API 的内核态调用处设置权限检查点,从而对应用的系统权限调用进行拦截.此方法的缺陷是需要对系统内核进行修补,此操作需要 root 权限,容易对系统造成威胁.并且,一旦用户对系统内核进行升级,则此系统补丁失效,需要重新根据新内核改写补丁并于更新,使用较为繁琐.

目前比较流行的权限管理解决方案是注入钩子[9].如目前市场占有率较高的 360 手机卫士和 LBE 安全大师.这些软件通常首先取得 Android 系统的 root 权限,接下来对系统框架层涉及到应用权限的系统调用进行挂钩.当一个应用需要进行权限调用时,会触发这些软件设置的钩子函数,通过软件内部的策略来决定权限调用是成功还是失败.这样的解决方案通用性较好,适用于几乎所有的系统版本.但是,其存在的问题是,需要进行 root 操作来获取最高权限,而 root 操作极易给系统带来隐患.

为了解决 Android 平台不同层次的权限限制和管理问题,本文提出了多层次的 Android 权限管理方法,在系统底层方面,提出了面向 MLS 的 Linux 用户权限模型,并给出了 Linux 用户权限关系树到 MLS 安全级定义的转换算法;在系统上层应用方面,提出了基于 Android 应用权限动态管理机制.首先设计安全策略文件的内容和格式,接着给出请求评估算法,并在系统调用关键位置设置权限检查点,从而有效解决 Android 应用权限的动态管理问题.

本文首先介绍目前现有的 Android 的权限控制方法,并分析其不足.第 1 节提出面向 MLS 的 Linux 用户权限模型.第 2 节提出上层应用的动态权限管理机制.第 3 节给出相关实现并进行实验验证.最后第 4 节总结全文,给出未来的研究方向.

## 1 底层的 Linux 用户权限模型

### 1.1 Linux用户权限关系树构建

为了解决新版本 Android 系统存在的 root 提权问题,本文提出基于 MLS 的 Linux 用户权限模型.Android 系统底层基于 Linux 内核,也具有 Linux 用户等概念.但是 Android 系统是单用户系统,Linux 传统的用户概念没

有意义,因此其在设计之初赋予了 Linux 用户不同的含义,将不同的 Linux 用户用来区分不同的应用或服务[10].Android 5.1.0 现在启用的有 52 个 Linux 用户,其中只有 15 个用户作为进程 UID 使用,其他 37 个用户只作为进程 GID 或者文件 UID/GID 来使用.这 15 个作为进程 UID 的用户分别是:

root,system,logd,nobody,shell,radio,drm,media,install,keystore,media_rw,camera,nfc,wifi,Bluetooth.

本文提出的 Linux 用户权限模型的核心思想就是给这 15 个用户采用树形结构进行权限大小的划分,当用户 $U_1$ 是用户 $U_2$ 的子节点时,我们有 $U_1 <_p U_2$,表示 $U_1$ 的权限被包含在 $U_2$ 的权限里.在这里,$<_p$ 是权限包含关系,p 意为"permission",该关系具有传递性.

当存在 3 个用户 $U_1, U_2, U_3$ 满足 $U_1 <_p U_2$ 和 $U_2 <_p U_3$ 时,可以自动得到 $U_1 <_p U_2$.

Android 系统虽然没有明确各 Linux 用户之间的关系(除了 root 权限最高外),但是本文通过分析 Android 源码的权限设置发现,其 Linux 用户之间的权限还是存在一些规律的,这些规律本文称其为用户权限规则.

**用户权限规则 I**. 通过 Android 源码中对 Linux 用户定义的说明,可以推断出用户之间的关系.

**用户权限规则 II**. 通过分析系统内置的文件、目录的属主和属组,根据一般属主比属组权限要高的原则,本文认为属主对应的 Linux 用户的权限包含属组对应的 Linux 用户的权限.

**用户权限规则 III**. 通过 Android 启动脚本中系统服务的属主和属组,根据一般属主比属组权限要高的原则,本文认为属主对应的 Linux 用户的权限包含属组对应的 Linux 用户的权限.

通过以上 3 个规则,本文分析出了 15 个作为进程 UID 的 Linux 用户之间的权限包含关系,见表 1.

**Table 1** Permission containment relationships of the Android's Linux users
**表 1** Android 系统中 Linux 用户之间的权限包含关系

| Android 代码位置 | 应用规则 | 权限包含关系 |
| --- | --- | --- |
| system/core/include/private/android_filesystem_config.h | 用户权限规则 I | 所有其他用户 $<_p$ root |
| system/core/rootdir/init.rc | 用户权限规则 II | radio $<_p$ system |
| system/core/include/private/android_filesystem_config.h | 用户权限规则 I | nobody $<_p$ system |
| system/core/include/private/android_filesystem_config.h | 用户权限规则 I | media_rw $<_p$ media |
| system/core/rootdir/init.rc | 用户权限规则 III | camera $<_p$ media |

通过将表 1 的权限包含关系进行分析整理,最终可形成 Linux 用户权限关系的树形结构,如图 1 所示.

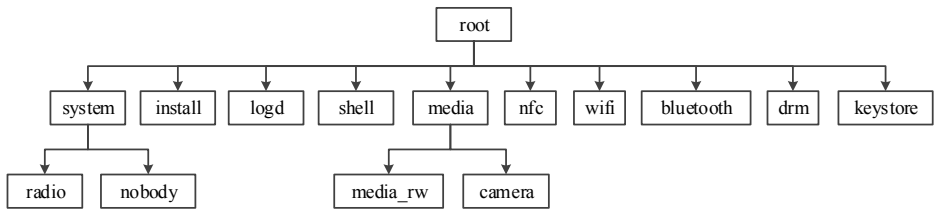

Fig.1 Relational tree of Linux user permissions
图 1 Linux 用户权限关系树

针对图 1 的 Linux 用户权限关系树,令 $U_1$ 和 $U_2$ 是两个 Linux 用户,可制定下列安全约束.

**约束 I**. 如果有 $U_1 <_p U_2$,则禁止从 $U_1$ 所属的进程对 $U_2$ 所拥有的文件进行读、执行类型的操作.

**约束 II**. 如果有 $U_1 <_p U_2$,则禁止从 $U_2$ 所属的进程对 $U_1$ 所拥有的文件进行写类型的操作.

**约束 III**. 如果有 $U_1 <_p U_2$,则禁止从 $U_1$ 所属的进程对 $U_2$ 所属的进程执行域转换.

其中,读类型的操作包括 getattr,read,ioctl,lock 等,执行类型的操作包括 execute,execute_no_trans,写类型的操作包括 append,write.

### 1.2 MLS 安全级的生成算法

Android 系统中的 MLS 机制的安全级是采用密级和范畴集来描述的,格式为 sensitivity[:category_set], sensitivity 为密级,目前 Android 系统只有一个密级 $s0$;category_set 是范畴集,当前 Android 系统的范畴集为空集合.由于 Android 本身并不直接支持我们在第 1.1 节提出的 Linux 用户权限模型,因此为了能够让该模型直接利

用 Android 现有的安全机制实现,本节提出一种 MLS 安全级的生成算法.该算法能够将第 1.1 节的 Linux 用户权限模型转换为采用密级和范畴集描述的 MLS 模型,并保证转换前与转换后的权限语义不变.

算法的核心思想是:密级分配按照从根节点到叶子节点的顺序进行分配,范畴集分配按照从叶子节点到根节点的顺序进行分配,分配完成后保证其 MLS 语义与 Linux 用户权限模型语义一致.具体的算法执行流程如下.

① 计算出 Linux 用户权限树的高度为 $n$,令树根节点的密级为 $s_{n-1}$;
② 若父节点的密级为 $s_i$,则其所有子节点的密级为 $s_{i-1}$;
③ 重复②的过程,直到树中所有节点都设置了密级;
④ 设置所有叶子节点的范畴集为 $\{N_m\}$,其中,$N_m$ 为叶子节点 $m$ 的用户名称;
⑤ 若子节点 $c$ 的范畴集为 $C_c$,父节点 $f$ 的范畴集为 $C_f$,则有 $C_f=C_c\{N_m\}$;
⑥ 重复⑤的过程,直到树中所有节点都设置了范畴集.

通过对第 1.1 节的 Linux 用户权限关系树执行上述算法,得到表 2 中的结果,其中密级 $s2>s1>s0$.

**Table 2** Generated MLS levels of the Linux users
**表 2** 生成的各 Linux 用户的 MLS 安全级

| Linux 用户 | MLS 安全级 |
|---|---|
| root | $s2$: {root,system,install,logd,shell,media,nfc,wifi,bluetooth,drm,keystore,radio,nobody,media_rw,camera} |
| system | $s1$: {system,radio,nobody} |
| install | $s1$: {install} |
| logd | $s1$: {logd} |
| shell | $s1$: {shell} |
| media | $s1$: {media,media_rw,camera} |
| nfc | $s1$: {nfc} |
| wifi | $s1$: {wifi} |

## 2 上层应用的动态权限管理模型

Android 平台中共有 130 多种权限[11],涵盖了电话、短信、网络、定位等诸多功能.在应用安装时,Android 终端用户可以查看到应用所申请的权限,用户可以直接在界面选择安装或者不安装.但是,Android 系统本身没有实现在应用程序运行时限制其权限的功能,因此,Android 用户无法对权限进行动态调整[12].有些应用由于开发人员疏忽,或者应用本身存在恶意企图,导致其会申请很多与自身无关的权限,威胁系统安全和用户数据的安全.如果有一种能够使终端用户在安装应用后对一部分其申请的权限进行动态调整和撤销的机制,那么 Android 的应用权限管理的灵活性将大大加强:终端用户可以根据其自身需求利用动态权限调整对其敏感的数据进行保护,或者利用安全软件的恶意代码指纹库等机制对已知恶意应用的权限进行屏蔽.从而增强 Android 的安全性.对此,本文提出了一种 Android 应用动态权限管理模型.其体系结构如图 2 所示.

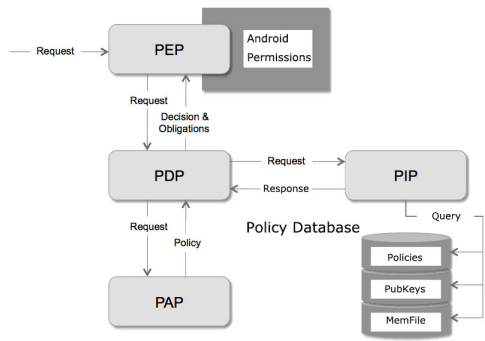

Fig.2　Dynamic management model of Android application permissions
图 2　Android 应用权限动态管理模型

策略决策点(policy decision point,简称 PDP)在本模型中处于核心的位置,其主要负责策略的决策[13].策略信息点(policy information point,简称 PIP)负责管理策略数据库(policy database),一方面向策略管理员(policy admission point,简称 PAP)提供查看,修改策略的接口,另一方面向 PDP 提供策略以用于决策.策略执行点(policy enforcement point,简称 PEP)主要负责向 PDP 传递请求,并执行 PDP 的访问控制决策.整体工作流程是:PDP 从 PEP 处接受请求,根据 PIP 处的策略评估该请求,并将授权决定返回给 PEP.

PIP 主要通过策略文件形式实现.下面我们给出策略文件的规格说明.

**2.1 策略文件规格说明**

策略文件位于策略数据库中,用来实现 Android 应用授权策略的存储,策略文件采用 XML 格式规范编写[14],下面给出其规格说明.

1. 签名是以十六进制编码的 X.509 证书.每个 signer 标签都必须包含签名.
2. ⟨signer signature=""⟩元素可以包含多个子元素.
   allow-permission:定义一系列许可的权限(白名单).
   deny-permission:定义禁止的权限(黑名单).
   allow-all:广义标签,给予所有权限.
   package:综合标签,为指定名称的受签名保护的包单独定义了允许、拒绝和广义子标签.
3. 允许定义 0 个或多个全局⟨package name=""⟩标签.这些标签可以为指定名称的包规定策略而不受签名的限制.
4. 当一个标签内有多个子元素时,将会采用下面的逻辑来决定采用什么策略.

如果定义了至少一个 deny-permission 标签,黑名单将会启用.

如果定义了至少一个 allow-permission 标签,并且不在黑名单内,白名单将会启用.

如果既没有黑名单也没有白名单,并且至少定义了一个 allow-all 标签,将会启用广义匹配策略(给予所有权限).

如果使用了⟨package name=""⟩子元素,该元素的权限控制策略将按照以上逻辑使用,并且会覆盖签名下的全局设置.

为了保证规则的正常使用,以上条件必须至少满足 1 项.

**2.2 动态请求评估算法**

在 Android 应用向系统申请权限调用时,PEP 会将相关访问请求发至 PDP,由 PDP 依据策略对该请求进行评估决策,返回允许或者拒绝.我们首先定义请求的格式.

request::={pkg_name,signature,perm}

其中,request 表示请求,pkg_name 表示申请权限的应用程序名称,signature 指代应用程序的签名,perm 是应用程序此次申请的权限名称.

根据第 2.1 节策略文件规格说明,我们可以实现 PDP 的评估决策算法,其伪代码如下所示.

```
1    INPUT: String pkg_name, Integer [] signature, String perm.
2    OUTPUT: Boolean permitted.
3    PROCEDURE checkPermission ()
4    String policy_file=file_open (POLICY_FILE_PATH);
5    Boolean whiteMode=get_mode (policy_file);
6    Integer pos_sig, pos_pkg, pos_perm;
7    pos_sig=search (policy_file, signature);
8    IF (pos_sig.exists=false)
9        RETURN (false);
10   pos_pkg=binary_search (policy_file, pkg_name, pos_sig);
11   IF (pos_pkg.exists=false)
12       RETURN (false);
13   pos_perm=binary_search (policy_file, perm, pos_pkg);
14   IF (pos_perm.exists=true AND whiteMode=true)
15       RETURN (true);
```

| | | |
|---|---|---|
| 16 | **ELSE IF** (pos_perm.exists=false **AND** whiteMode=false) | |
| 17 |     **RETURN** (true); | |
| 18 | **ELSE** | |
| 19 |     **RETURN** (false); | |
| 20 | **END** checkPermssion | |

假设 Android 系统中共有 $n$ 个应用程序,平均每个应用程序申请 $m$ 项权限,上述的请求评估算法的时间复杂度为 $O(mn)$.

## 3 实现及实验验证

本文方法所采用的硬件环境为友坚恒天 UT4412 开发板和 Google Nexus 5 手机,其中,UT4412 开发板搭载了 Android 4.4 操作系统,Nexus 5 手机则搭载了 Android 5.1 操作系统,这两种产品在 Android 平板和手机领域具有一定的代表性,因此本文方法在两个系统上都进行了实现.其中,底层的 Linux 用户权限模型主要在内核层借助 MLS 机制实现,而上层应用的动态权限管理机制则在 Android 框架层实现.下面分别予以详细阐述.

Linux 用户权限模型主要利用 Android 中的 MLS 机制进行实现.具体实现主要分为两个步骤.

(1) 首先依据第 1.1 节中提出的 Linux 用户权限模型的 3 条安全约束编写相应的 MLS 规则,如下所示.

• mlsconstrain {file} {getattr, read, ioctl, lock, execute, execute_no_trans} (l1 dom l2);

• mlsconstrain {file} {append, write} (l1 domby l2);

• mlsconstrain process {transition} (l1 dom l2).

为了使这 3 条 MLS 规则生效,需要将其加入到 MLS 的约束配置文件 mls 中,其位于源码的 external/sepolicy 目录下.

(2) 然后依据第 1.2 节中提出的 MLS 安全级生成算法,将 Linux 用户权限关系树中的每一个 Linux 用户都赋予相应的 MLS 安全级,由于 Android 中的 MLS 机制是与 SELinux 捆绑在一起实现的,因此这部分需要修改 SELinux 的标签初始化文件,如服务的标签文件 service_contexts、文件、目录的标签文件 file_contexts、属性的标签文件 property_contexts,将这些文件内定义的安全上下文中默认的 $s0$ 安全级替换为本文表 2 中定义的安全级.

根据第 2 节,Android 上层应用的动态权限管理模型需要实现 PEP,PDP,PAP 和 PIP 等部分.请求评估算法通过框架层来实现,策略文件的路径为"/data/pekisafe/mac_permissions.xml",在目标权限的调用路径加入权限检查点后,调用请求评估算法进行授权检查,并返回授权结果.

PEP 负责执行决策,主要由权限检查代码构成.Android 系统的权限较多,且调用路径较为分散.因此,需要将权限检查代码安插在所有需要监控的权限调用路径上,在成功调用权限之前,先进行权限检查.若权限检查返回结果为 allow,则延续原有调用路径;若权限检查返回 deny,则通过空指针或者 return 返回,中断原有调用路径.

由于权限检查需要拦截 Android 原有的系统调用,并进行策略文件比对来进行决策,因此势必会对系统性能造成一定的影响,为了不对过多的函数调用造成影响,本方法检查点位置的选取以尽量靠近更高的层次为原则.通过分析我们发现,在 Android 框架层加入检查点比较合理.框架层检查点不仅可拦截所有应用层次的调用,同时也不会对系统调用产生过大影响.

Android 系统中几个主要权限检查点位置及其说明,见表 3.

为了验证 Linux 用户权限模型的防护效果,我们针对市面上 5 种流行的 root 工具进行了测试,其中 KingRoot 是应用类型 root 工具,其他 4 款都是通过 PC 端进行操作的工具.实验结果见表 4.

实验结果表明,只有 KingRoot 和刷机大师能够安装授权管理器和 su 模块,KingRoot 因为其是 APP 类型的 root 工具,为了测试其功能,我们特意将其加入到白名单中.刷机大师则是利用临时 root 权限突破了拦截实现了自身 APK 应用包的安装.其他 3 种工具的 APK 安装都被安装权限管理机制所拦截,从而导致 root 失败.尽管 KingRoot 和刷机大师实现了 su 的安装,但是两者都无法以 root 权限执行功能,如利用 Root Explorer(一个利用 root 权限进行文件管理的软件)打开系统所保护的核心文件.因此可以认为我们的保护是有效的.

为了验证应用动态权限管理机制的可用性,本文选取了 50 个 Android 应用在本平台进行测试.为了保证实验结果的客观性,实验样本分为两部分:其中 30 个样本来源自各大 Android 应用平台,如 Google Play,安卓市场等,另外 20 个样本来自于专门的恶意软件样本库[15],50 个应用各取一个核心权限进行拦截.经测试,本方法权限拦截准确率是 98%.唯一失败的测试用例是百度地图,在撤销 GPS 权限后,仍然可以进行定位和使用.经分析,这是由于百度地图集成了 GPS 定位、蜂窝网络定位、WIFI 基站定位 3 种定位方式,即使撤销其 GPS 权限,只要其还接入了通信网络或 WiFi 网络,就仍可进行定位.经过对蜂窝网络权限和 WiFi 权限进行撤销后,我们也实现了百度地图权限的正常拦截.部分样本的测试结果见表 5.默认情况下所有权限一开始都是授予状态的,当用户察觉到恶意应用,如百度地图有读取短信行为后,则可以打开本方法所提供的权限管理页面,对瓦力短信的"读取 SMS 权限"进行撤销,从而杜绝该应用类似行为的再次发生.

**Table 3** Check points of Android permissons
**表 3** Android 权限检查点

| 权限 | 检查点所在源文件 | 检查点说明 |
|---|---|---|
| android.permission.READ_CONTACTS (读取联系人) | frameworks/base/core/java/android/content/ContentResolver.java, Line 303 | query 函数中添加 String 比较代码,比较对象为"com.android.contacts"和"contacts" |
| android.permission.WRITE_CONTACTS (写入联系人) | frameworks/base/core/java/android/content/ContentResolver.java, Line 759 | applyBatch 函数中添加代码,不需要比较 |
| android.permission.READ_SMS (读取短信) | frameworks/base/core/java/android/content/ContentResolver.java, Line 303 | query 函数中添加 String 比较代码,比较对象为"sms" |
| android.permission.SEND_SMS (发送短信) | frameworks/base/telephony/java/android/telephony/SmsManager.java, Line 226 | getDefault 函数中添加代码,不需要比较 |
| android.permission.RECORD_AUDIO (录音) | frameworks/base/media/java/android/media/MediaRecorder.java, Line 97 | MediaRecorder 函数中添加代码,不需要比较 |
| android.permission.CAMERA (拍照) | frameworks/base/core/java/android/hardware/Camera.java, Line 263, 267 | open 函数中添加代码,不需要比较 |
| android.permission.ACCESS_FINE_LOCATION (定位) | anframeworks/base/core/java/android/app/ContextImpl.java, Line 1174 | getSystemService 函数中添加 String 比较代码,比较对象为"location" |
| | frameworks/base/location/java/android/location/LocationManager.java, Line 992 | isProviderEnabled 函数中添加代码,不需要比较 |
| | frameworks/base/location/java/android/location/LocationManager.java, Line 1019 | getLastKnownLocation 函数中添加代码,不需要比较 |
| | frameworks/base/location/java/android/location/LocationManager.java, Line 566 | _requestLocationUpdates 函数中添加代码,不需要比较 |

**Table 4** Privilege escalation results of the five popular root tools
**表 4** 5 种流行 root 工具的提权结果

| Root 工具 | 版本 | 工具类型 | 权限提升结果 |
|---|---|---|---|
| 360 一键 Root | 5.3.3.0 | PC | root 失败,无法安装应用 APK 包 |
| KingRoot | 4.0 | APP | 能够安装授权管理器和 su 模块,但是无法以 root 权限运行应用 |
| 百度一键 ROOT | 3.5.09 | PC | root 失败,无法安装应用 APK 包 |
| Root 精灵 | 2.2.7 | PC | root 失败,无法安装应用 APK 包 |
| 刷机大师 | 4.1.1 | PC | 能够安装其 APK 包和 su 模块,但是无法以 root 权限运行应用 |

**Table 5** Permission revocation results for several apps
**表 5** 部分应用权限撤销对运行造成的影响

| 应用名称 | 权限定义 | 权限描述 | 撤销后影响 |
|---|---|---|---|
| com.fci.easyContacts 1.0.6 (通话录) | android.permission.READ_CONTACTS<br>android.permission.WRITE_CONTACTS | 读取联系人权限<br>写入联系人权限 | 应用崩溃<br>联系人添加无效 |
| cn.com.wali.walisms 3.8.1 (瓦力短信) | android.permission.READ_SMS<br>android.permission.SEND_SMS | 读取 SMS 权限<br>发布 SMS 权限 | 应用无法显示短信<br>短信发送无效 |
| com.gui.gui.chen.soundrecorder 1.2.3 (Sound Recorder) | android.permission.RECORD_AUDIO | 录音权限 | 应用崩溃 |
| org.winplus.camera 1.0 (CameraDemo) | android.permission.CAMERA | 拍照权限 | 应用崩溃 |
| com.tencent.map 4.5.0 (腾讯地图) | android.permission.ACCESS_FINE_LOCATION | 定位权限 | 无法接收到定位信号 |

从表 5 中可以看出,一部分应用在权限被撤销后发生崩溃,如读取联系人权限、录音权限和拍照权限.需要

说明的是,发生崩溃是由于应用自身在异常处理机制不够完善所导致的,Android 本身也并无有效机制防止崩溃,因此问题还需要应用发行者自身来解决.其他应用在权限撤销后,相应权限的功能均被有效禁用.

同时,我们测试了在允许权限的情况下,权限检查点对应用运行效率所带来的影响.为了保证测试结果的精确,本文通过 Java 代码实现了上述表 5 中的权限调用功能,并在调用前后利用微秒级时间戳 API 来统计时间开销.统计结果如图 3 所示.

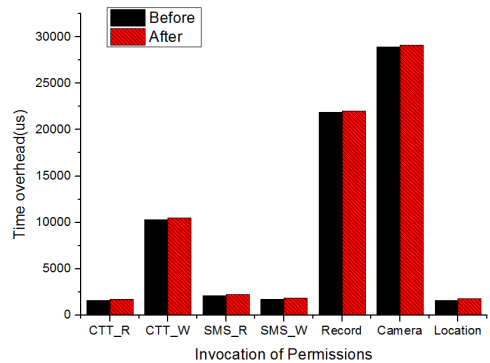

Fig.3　Effect of permission checking points to running efficiency
图 3　权限检查点对应用运行效率的影响

由图 3 中可以看出,检查点代码所造成的时间开销比较小,相比未插入权限检查代码时平均只增加了 140us 左右,而时间开销最小的读取联系人权限也至少需要 1 732us.相比之下,权限检查代码所造成的时间开销几乎可以忽略不计.因此可见,权限检查所增加的时间开销并不影响 Android 系统性能以及用户体验,是可以接受的.

## 4　结束语

本文首先指出了 Android 平台的两大安全威胁:(1) 安全漏洞频发;(2) 权限控制机制不健全,接着着重分析了 Android 底层机制和上层应用在权限控制方面存在的问题,进而提出了一种面向 Android 的多层次权限控制方法.该机制首先利用策略文件对 Android 应用授权进行语义描述,然后在 Android 系统权限调用路径进行代码插桩,通过请求评估算法来保证应用授权的有效评估.实验结果表明,UT4412 开发板以及 Nexus 5 手机经过安全加固后,能够有效阻止恶意软件的入侵及滥用权限等行为,同时不会显著增加系统开销,从而不影响用户体验.下一步工作将会考虑利用 SELinux 等内核安全机制对安全策略文件本身及其权限调用路径进行保护,进一步提高系统安全性.

**References**:

[1] Arzt S, Bartel A, Gay R, *et al*. Poster: Software security for mobile devices. In: Proc. of the 36th IEEE Symp. on Security and Privacy. Piscataway: IEEE, 2015. 1–2.

[2] Enck W, Ongtang M, Mcdaniel P. Understanding android security. In: Proc. of the 30th IEEE Symp. on Security and Privacy. Piscataway: IEEE, 2009. 50–57.

[3] Zhang YQ, Wang K, Yang H, *et al*. Survey of Android OS security. Journal of Computer Research and Development, 2014,51(7): 1385–1396 (in Chinese with English abstract).

[4] Lei LG, Jing JW, Wang YW, *et al*. A behavior-based system resources access control scheme for Android. Journal of Computer Research and Development, 2014,51(5):1028–1038 (in Chinese with English abstract).

[5] Yang H, Zhang YQ, Hu YP, *et al*. A malware behavior detection system of Android applications based on multi-class features. Chinese Journal of Computers, 2014,37(1):15–27 (in Chinese with English abstract).

[6] Grace M, Zhou Y, Zhang Q, *et al*. Riskranker: Scalable and accurate zero-day Android malware detection. In: Proc. of the 10th Int'l Conf. on Mobile Systems, Applications, and Services. New York: ACM, 2012. 281–294.

[7] Wang HY, Wang ZY, Guo Y, *et al*. Detecting repackaged Android applications based on code clone detection technique. SCIENTIA SINICA Informationis, 2014,44(1):142−157 (in Chinese with English abstract).

[8] Sarwar G, Mehani O, Boreli R, *et al*. On the effectiveness of dynamic taint analysis for protecting against private information leaks on Android-based devices. In: Proc. of the 10th Int'l Conf. on Security and Cryptography. Berlin, Heidelberg: Springer-Verlag, 2013. 461−468.

[9] Conti M, Nguyen VTN, Crispo B. CRePE: Context-Related policy enforcement for Android. In: Proc. of the 13th Information Security Conf. Berlin, Heidelberg: Springer-Verlag, 2011. 331−345.

[10] Tesfay WB, Booth T, Andersson K. Reputation based security model for Android applications. In: Proc. of the 11th Int'l Conf. on Trust, Security and Privacy in Computing and Communications (TrustCom). Piscataway: IEEE, 2012. 896−901.

[11] Zhang Y, Yang M, Xu B, *et al*. Vetting undesirable behaviors in Android apps with permission use analysis. In: Proc. of the 2013 ACM SIGSAC Conf. on Computer & Communications Security. New York: ACM, 2013. 611−622.

[12] Felt AP, Ha E, Egelman S, *et al*. Android permissions: User attention, comprehension, and behavior. In: Proc. of the 8th Symp. on Usable Privacy and Security. New York: ACM, 2012. 3−16.

[13] Standard O. Extensible access control markup language (xacml) version 2.0. 2005.

[14] Wang YZ, Feng DG, Zhang LW, Zhang M. XACMl policy evaluation engine based on multi-level optimization technology. Ruan Jian Xue Bao/Journal of Software, 2011,22(2):323−338 (in Chinese with English abstract). http://www.jos.org.cn/1000-9825/3707.htm [doi: 10.3724/SP.J.1001.2011.03707]

[15] Zhou Y, Jiang X. Dissecting Android malware: Characterization and evolution. In: Proc. of the 33th IEEE Symp. on Security and Privacy. Piscataway: IEEE, 2012. 95−109.

**附中文参考文献**:

[3] 张玉清,王凯,杨欢,方喆君,王志强,曹琛.Android 安全综述.计算机研究与发展,2014,51(7):1385–1396.

[4] 雷灵光,荆继武,王跃武,张中文.一种基于行为的 Android 系统资源访问控制方案.计算机研究与发展,2014,51(5):1028–1038.

[5] 杨欢,张玉清,胡予濮,刘奇旭.基于多类特征的 Android 应用恶意行为检测系统.计算机学报,2014,37(1):15–27.

[7] 王浩宇,王仲禹,郭耀,陈向群.基于代码克隆检测技术的 Android 应用重打包检测.中国科学(信息科学),2014,44(1):142–157.

[14] 王雅哲,冯登国,张立武,张敏.基于多层次优化技术的 XACML 策略评估引擎.软件学报,2011,22(2):323–338. http://www.jos.org.cn/1000-9825/3707.htm [doi: 10.3724/SP.J.1001.2011.03707]

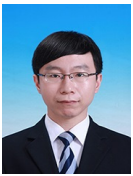
**罗杨**(1989−),男,河北衡水人,博士生,CCF 学生会员,主要研究领域为移动终端安全,云计算安全.

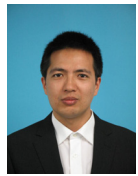
**刘宏志**(1982−),男,博士,副教授,CCF 会员,主要研究领域为计算机图形学,人机交互,云计算.

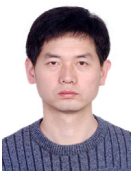
**张齐勋**(1979−),男,博士生,讲师,CCF 学生会员,主要研究领域为云计算,嵌入式软件工程.

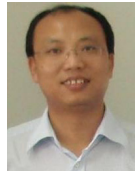
**吴中海**(1968−),男,博士,教授,博士生导师,CCF 高级会员,主要研究领域为情境感知服务,云安全与隐私保护,嵌入式智能,大数据与信息融合.

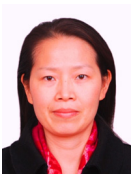
**沈晴霓**(1970−),女,博士,教授,博士生导师,CCF 高级会员,主要研究领域为操作系统与虚拟化安全,云计算安全,大数据安全与隐私,可信计算.